\documentclass[aps,prl,showpacs,twocolumn]{revtex4}

\def\beq{\begin{equation}}
\def\eeq{\end{equation}}
\def\beqn{\begin{eqnarray}}
\def\eeqn{\end{eqnarray}}
\def\bea{\begin{eqnarray}}
\def\eea{\end{eqnarray}}
\def\be{\begin{equation}}
\def\ee{\end{equation}}

\begin{document}

\voffset 1.25cm

\title{The Sterile Neutrino: First Hint of 4th Generation Fermions?}
\author{Stephen Godfrey and Shouhua Zhu}
\affiliation{Ottawa-Carleton Institute for Physics, \\
Department of Physics, Carleton University, Ottawa, Canada K1S 5B6}

\date{\today}

\begin{abstract}

In this letter, we introduce the ``flipped see-saw mechanism'',
a new type of see-saw mechanism with 
4th-generation neutrinos. This mechanism naturally explains
the light sterile neutrino which is needed to
account for all neutrino oscillation data. 
At the same time it predicts that another Majorona neutrino should 
exist with mass of the electro-weak scale. 
We comment on some implications of this scenario on the
oblique parameters used to parameterize precision electroweak 
measurements as well as on future experiments.

\end{abstract}

\maketitle

Over the past several years 
our understanding of neutrino physics has undergone important advances.
The combined results
from atmospheric neutrino measurements, solar neutrino measurements
and the long baseline experiments 
\cite{Review} imply the neutrino mass differences
$1.2 \times 10^{-3} < \Delta m_{23}^2 < 4.8 \times
10^{-3}$~eV$^2$ and $5.4 \times 10^{-5} < \Delta m_{12}^2 < 9.5
\times 10^{-5}$~eV$^2$.
However, there exists a serious problem in that the 
Los Alamos
Liquid Scintillator Neutrino Detector
(LSND) experiment \cite{LSND}
finds $10 > \Delta m^2 >0.2$~eV$^2$ which is in 
serious conflict with the other results.  It is possible that the LSND 
result is in error.  The Mini-Boone experiment at Fermilab \cite{Bazarko:1999hq}
is studying the appropriate region of parameter space and will be able to 
either confirm or rule out the LSND result.  However,  accepting
the LSND result along with the limit on 3 light neutrino 
species from LEP-SLC measurements of $Z^0$ decay
implies the need for a sterile 
neutrino which has little or no interaction with the $W$ 
and $Z$ bosons \cite{OneSterileNeutrino,TwoSterileNeutrino}
(notwithstanding that a
sterile neutrino poses a challenge to standard big bang 
nucleosynthesis calculations \cite{Cirelli}).
SNO neutral current data implies that  $\nu_\mu  \rightarrow \nu_e$
must proceed via $ \nu_\mu \rightarrow
\nu_s \rightarrow \nu_e$ \cite{Review} and while ruling
out the pure $\nu_e \rightarrow \nu_s$
transition allows as much as 40\% admixture of
sterile neutrino \cite{Review}.
For the most part models of light sterile neutrinos have been 
introduced on purely phenomenological grounds to accommodate the LSND 
result in fits of neutrino mixing data without explaining it's origins.
In this paper we describe a 
mechanism which naturally explains a light sterile neutrino.

It is well known that the tiny mass of neutrinos can be related to
the high mass scale of the right-handed Majorona neutrino via the see-saw
mechanism which offers a natural explanation for the light
SM neutrinos. The conventional see saw mechanism
does not explain why the sterile neutrino
mass is so small.  Some explanations are
that the small sterile neutrino mass 
is related to the mechanism of supersymmetry breaking
\cite{Babu:2000hb}, that it is protected by a gauge symmetry 
\cite{Babu:2003is}, or by the introduction of yet more neutrinos
in conjunction with a double see-saw mechanism \cite{McDonald:2004pa}.
 In this note we propose the flipped see-saw
mechanism which can naturally induce the small sterile
neutrino mass. The idea is based on a natural extension of the
SM that adds a 4th generation of fermions \cite{Hill:1989vn,Fargion:1999ss}.
The scenario we are proposing differs from the
well known Hill-Paschos scenario \cite{Hill:1989vn}
in that it induces a light right-handed Majorona neutrino
and an electro-weak scale Majorona neutrino
whereas the two Majorona neutrinos 
of the Hill-Paschos scenario both have electro-weak scale masses.

Heavy 4th generation chiral fermions are disfavoured by 
precision observables if one considers
the contributions to the oblique parameters \cite{Peskin:1990zt}
$S$ and $T$ separately. 
The contributions to
oblique parameters from possible new physics contributions have
been classified into 3 cases \cite{Peskin:2001rw}: (1) decreasing
$S$, (2) increasing $T$, and (3) achieving both. 
The authors of Ref. \cite{He:2001tp},
argued that the 4th generation SM-like fermions
fall into case (2) and are consistent with precision measurements
without requiring new physics. 
We analyze the oblique precision constraints on our scenario below.

The 4th generation neutrino scenario we propose is rather straight 
forward. We refer to it as the ``flipped see-saw mechanism''. 
For simplicity we omit mixing
with the other 3 generations.  The light neutrino is then 
obtained via the mass matrix 
\begin{equation}
{1\over 2}
\overline{\omega^c}
\left( 
\begin{array}{cc}
M & D \\
D & 0 
\end{array}  
\right)
\omega
\end{equation}
with
\begin{equation}
\omega=\left(\begin{array}{c}
\nu_L^c \\
\nu_R
\end{array} 
\right),
\end{equation}
where $\nu_L$ and $\nu_R$ are the left- and right-handed neutrinos of
the fourth generation and  
$\nu_L^c\equiv C (\bar{\nu}_L)^T$ is the (right-handed) charge
conjugate field.
Here $D$ represents the usual Dirac mass term and $M$ the 
left-handed  Majorona mass term which is 0 in the normal see-saw 
mechanism.  $M$ can be induced via new dynamics. For example, $SU(2)$ 
Higgs triplet fields $\phi$ via \cite{Gelmini:1980re}, 
\be 
L=-G_\nu \overline {L^c} \tau . \phi L
\ee 
with $L=(\nu_L, \ell_L)$.  In this scenario
the triplet vacuum expectation value is constrained by electroweak 
fits of the $\rho$ parameter to be small implying a large 
non-perturbative Yukawa coupling which we assume is due to new unknown 
physics.
It should be emphasized that, the electro-weak scale, $M$, 
may induce some difficulties in triplet models, so that
the origin of $M$ might 
due to other unkown dynamics.
The right-handed Majorona
mass term is 0 in contrast to the large
mass of the usual Majorano model which can be induced via extra
symmetry for the right-handed neutrino.

The neutrino masses are obtained by diagonalizing the mass matrix
and a small neutrino mass is obtained if $M \gg D$ and is given by
\be 
m_\nu \sim \frac{D^2}{M}
\ee 
with mixing
angle $\theta =D/M$. So for example, for $M=200$ GeV and $D=0.3$ MeV, $m_\nu
=0.9$ eV, which is in the range of the LSND result.  
This Majorona neutrino can
easily escape the constraint from $Z^0$ decay data because of the
$\theta^2$ suppression factor for left-handed couplings.  The heavy 
neutrino mass is $m_N \sim M$, the electroweak scale.

This scenario differs from that of Hill and Paschos \cite{Hill:1989vn}. 
In the 
Hill-Paschos scenario if the Majorona mass term for right-handed
neutrino,  $M$, is the electro-weak scale then $D$ must be of the same order 
as $M$ or else the induced light Majorona neutrino will show up in
$Z^0$ decay. In the flipped see-saw mechanism the light Majorona 
neutrino almost totally decouples from the 
electro-weak interaction. The heavy Majorona neutrino mass is of the 
order of the 
electro-weak scale and should contribute to the oblique
parameters $S$, $T$ and $U$. 

We next consider the constraints we can put on 
4th generation fermions using
the Peskin-Takeuchi \cite{Peskin:1990zt} parameterization of the 
oblique corrections, $S$, $T$ and $U$.  Our definitions 
for  $S$ and $U$  are slightly different from the original Peskin 
Takeuchi definitions in that we use the 
differences of the $\Pi$ rather than their first derivatives which 
has the benefit of eliminating the mass singularity.  The oblique 
corrections are given by  \cite{Peskin:1990zt}:
\bea S
&=& -16 \pi \frac{\Pi_{3Y}(m_Z^2)-\Pi_{3Y}(0)}{m_Z^2} \nonumber \\
T &=& 4 \pi \frac{\Pi_{11}(0)-\Pi_{33}(0)}{x_w (1-x_w) m_Z^2}
\nonumber \\
U &=& 16 \pi \frac{[\Pi_{11}(m_Z^2)-\Pi_{11}(0)]-
[\Pi_{33}(m_Z^2)-\Pi_{33}(0)]}{m_Z^2} 
\eea 
where $\Pi_{11}$ and
$\Pi_{33}$ are the vacuum polarizations of isospin currents, and
$\Pi_{3Y}$ is the vacuum polarization of one isospin and one
hyper-charge current, and $x_W=\sin^2\theta_W$ with $\theta_W$
weak angle defined at $m_Z$. 

The contributions to $T$ from the 4th generation leptonic sector 
in the limit $m_\nu \rightarrow 0$ and $\theta \rightarrow 0$
can be written \cite{Kniehl:1992ez}:
\be 
\Delta T=\frac{1}{16 \pi x_W} \frac{m_E^2}{m_W^2}\label{teq}
\ee 
where $m_E$ is the 4th generation charged lepton mass.
The weakest constraint on $m_E$ is obtained when there is no quark 
contribution to $\Delta T$ which occurs for the case of degenerate 
4th generation $U$ and $D$-type quarks.  
Using the most conservative value of $\Delta T <0.6$ \cite{He:2001tp}
results in the limit of 
\be m_E < 210\ GeV.
\ee  
For comparison the direct limit from LEP of $m_E > 100$~GeV 
would imply $\Delta T > 0.14$.

We can likewise use $S$ and 
$U$ to put constraints on 4th generation fermions where $m_N$ and 
$m_E$ enter as the masses of the lepton weak isospin doublet.
$\Delta S$ and $\Delta U$ can be written as  \cite{Kniehl:1992ez}:
\bea \Delta S &=&
\frac{\pi}{m_Z^2} \left\{
-2 [Re \Pi^A(m_Z^2,m_N,m_N)-\Pi^A(0,m_N,m_N)] \right. \nonumber \\
&&  + 3 Re \Pi^V(m_Z^2,m_E,m_E)- Re \Pi^A(m_Z^2,m_E,m_E)
\nonumber \\
&& \left. +\Pi^A(0,m_E,m_E)\right\} \\
\Delta U&=& 2 \pi \left\{ -\frac{1}{m_W^2} \left[
Re(\Pi^V+\Pi^A)(m_W^2,m_N,m_E)\right. \right.\nonumber \\
&& \left.- Re (\Pi^V+\Pi^A)(0,m_N,m_E)\right] \nonumber \\
&&\left.+\frac{2}{m_Z^2} Re \Pi^V(m_Z^2,m_E,m_E) \right\}-\Delta
S, \eea where \bea \Pi^{V,A}(q^2,m_1,m_2) &=& \frac{1}{12 \pi^2}
\left\{ \left[ q^2-\frac{m_1^2+m_2^2}{2}\pm 3 m_1 m_2 \right.
\right.
\nonumber \\
&& \left. -\frac{(m_1^2-m_2^2)^2}{2 q^2} \right]
B_0(q^2,m_1^2,m_2^2) \nonumber \\
&&+ m_1^2 \left[ -1+\frac{m_1^2-m_2^2}{2 q^2}\right]
B_0(0,m_1^2,m_1^2) \nonumber \\
&&+ m_2^2 \left[ -1+\frac{m_2^2-m_1^2}{2 q^2}\right]
B_0(0,m_2^2,m_2^2) \nonumber \\
&& \left. -\frac{q^2}{3}+\frac{(m_1^2-m_2^2)^2}{2 q^2}\right\}
\label{eqk} \eea with $B_0$ the standard scalar loop two-point
function also given in ref. \cite{Kniehl:1992ez}.

\begin{figure}[thb]
\vbox{\kern2.8in\includegraphics{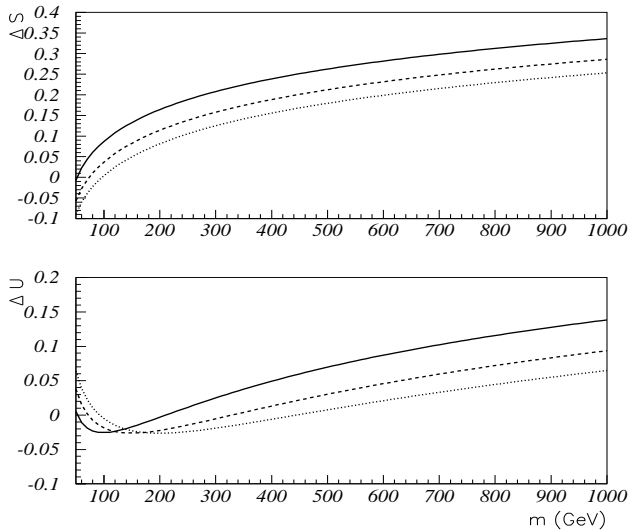}}
\caption{ $\Delta S $
and $\Delta U$ as a function of neutrino mass for $m_E=$100
[solid], 150 [dashed] and 200 GeV [dotted].}
\end{figure}

Fig. 1 shows $\Delta S$ and $\Delta U$ plotted as a function of the
heavy Majorona neutrino mass, $m_N$, for $m_E=100$, 150, and 200 GeV. 
While one can see that the present $U$ measurement of $U=0.18\pm 0.14
(+0.01)$  \cite{He:2001tp} \cite{higgs}
 does not constrain the
neutrino mass, the $S$ parameter value of 
$S=-0.04\pm 0.11 (0.09)$  constrains $m_N$ to not 
be very large.  This is especially true 
for heavy 4th generation quarks where degenerate chiral quarks contribute 
$\Delta S^q = 1/2\pi\approx 0.16$ in addition to a
TeV Higgs boson which contributes $\Delta S \sim 0.15$ relative to the 
reference Higgs mass of 100~GeV \cite{higgs2}.  
From these considerations it is 
unlikely that both heavy quarks and a heavy Higgs boson (say $M_H>400$ GeV)
exist.  Likewise, the contributions to $\Delta S$ from
heavy 4th generation quarks 
or a heavy Higgs boson 
leaves no room for further contributions from a heavy neutrino.
However, a light neutrino, say less than 100
GeV for $m_E=200$, can make a negative contribution to $S$
\cite{Gates:1991uu}
which would partially cancel the positive contributions from 
heavy quarks.  
It is therefore possible for the scenario of a 4th generation 
lepton and neutrino with $ m \sim {\cal O}(100)$~GeV
and 4th generation quarks  to survive.

If we assume heavy, degenerate, 4th generation quarks and 
$M_H=100$~GeV we obtain the allowed parameter space for $m_E$ and 
$m_N$ shown in Fig. 2 using the $S$ and $T$ values from Ref. 
\cite{He:2001tp} and taking $U=0$.  As the Higgs mass increases the 
allowed region shrinks, mainly due to the Higgs contributions to $S$.

\begin{figure}[thb]
\vbox{\kern2.8in\includegraphics{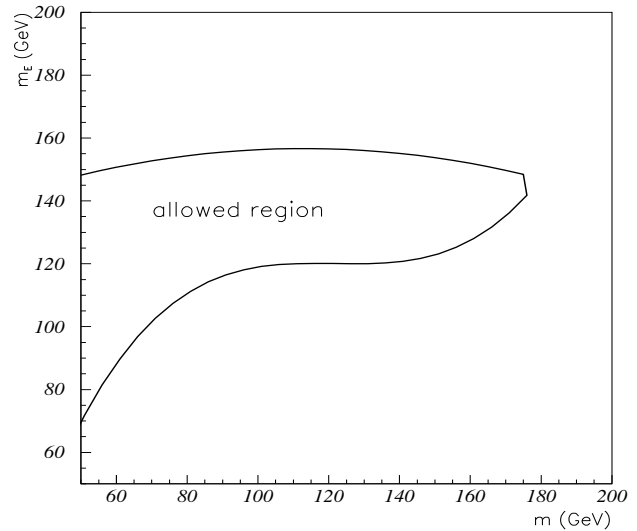}} 
\caption{ 95\% CL
allowed region on neutrino($m$) and charged-lepton ($m_E$) plane.
Here the S, T values are taken from Ref. \cite{He:2001tp}
 for
$U=0$ and $m_H=100$ GeV. The heavy quark contribution to T is set
to 0 and $S=1/2 \pi$ [degenerate quark masses]. }
\end{figure}

In this note we proposed a ``flipped see-saw mechanism'' for
4th generation neutrinos which naturally induces the light sterile 
Majorana neutrino needed to explain neutrino mixing 
measurements, in particular the LSND results.  
Constraints from existing high precision electroweak data implies a 
4th generation of fermions which consists of
\begin{enumerate} 
\item Two Majarona neutrinos, one which is nearly right-handed and
light ($\le 1$ eV) to account for the current
neutrino data, and another which is nearly left-handed and with mass
$M_Z/2 < M_N < 200$~GeV.
\item One charged lepton with $100 < M_E < 200$~GeV.
\item Two heavy (nearly) degenerate quarks with mass $\ge 200$ 
constrained by Tevatron search limits \cite{Limit}.
\end{enumerate}

In addition to constraints from oblique parameters, 
4th generation fermions can
also affect other low energy processes such as $\bar{B}^0-B^0$ and
$\bar K^0-K^0$ mixing, $b \rightarrow s \gamma$ etc. However, such
effects are significantly suppressed by the small mixing between SM and
4th generation fermions. In general, the constraints
on 3rd-4th generation mixing are looser than those on 1st-4th and
2nd-4th generation mixing. 
Constraints for mixing between 4th generation 
and SM fermions is reviewed in Ref. \cite{Frampton:1999xi}.

The key ingredient of this scenario will be tested in the near future
by the Mini-Boone experiment \cite{Bazarko:1999hq} which is searching
for the appearance of an electron neutrino from muon neutrino in the 
process $\bar\nu_\mu \rightarrow \bar \nu_e$.  If 
$\bar\nu_\mu \rightarrow \bar \nu_e$ is not observed the model must be 
either improved or discarded.
The remaining 4th generation
fermions can be directly produced at both hadron and $e^+e^-$ 
colliders provided it is kinematically allowed.  The
search strategies will depend on the fermion mass and 
the size of the mixing with SM fermions.
For example, in the lepton sector if the
3rd and 4th generation mixing angle, $\theta_{34}$, is not very
small, then both $N$ and $E$ will 
decay via $N\rightarrow \tau W^*$ and $E \rightarrow \nu_\tau W^*$. 
If the angle is very small
(of the order of $m_{\nu_\tau}/m_N$ or less), then the heavier one
(either $E$ or $N$) will decay into the lighter, while the lighter one
decays to SM normal fermions with a greatly suppressed
rate\cite{Frampton:1999xi}.  The situation is similar in the quark sector
and the phenomenology  of 4th generations 
can be found in Ref. \cite{Frampton:1999xi}. However
a unique feature of  the flipped see-saw model which should be 
pointed out 
is that the Higgs-neutrino coupling is proportional to 
$\sqrt{m_\nu m_N}/m_W$ in contrast to 
$m_N/m_W$ in conventional 4th generation models \cite{Novikov:2001md}.
This feature provides
one possible method to distinguish these two models.

The effects of 4th generation fermions will also appear via
virtual loops. The $ggH$ vertex  is especially sensitive
to heavy quarks which leads to an enhancement in the  $gg \rightarrow H$
production rate which can be as large as a factor of $\sim 10$.
Such effects might/must show up
at upgraded Tevatron/LHC \cite{Arik:2003cy}. The
enhancement due to heavy quarks will also occur in the
$\gamma\gamma H$ vertex, which will impact, for example, the Higgs
production rate at a $\gamma \gamma$ collider, as well as the
intermediate mass Higgs searches at the LHC via $H \rightarrow \gamma
\gamma$.

If the LSND result stands up to the scrutiny of the Mini-Boone 
experiment it may give the first evidence for the existence for a 
4th generation of fermions.


\noindent
{\em Acknowledgements}: 
The authors thank J. Erler, G. Marandella, A. Mastrano, and R. Volkas
for helpful communications.
This work was supported in part by the
Natural Sciences and Engineering Research Council of Canada


\begin{thebibliography}{99}
\bibitem{higgs}
The number of parenthesis is due to the
change of Higgs mass from 100 GeV to 300 GeV.
\bibitem{higgs2}
This result does not change significantly by using the LEP limit of 114~GeV.
\end{thebibliography}
\end{document}